# STUDY ON THE HYDROGEN BEHAVIOR IN Y AND Co DOPED BARIUM-ZIRCONATE

M. Khalid Hossain*[1,2], Kenichi Hashizume**[1]

1. Department of Advanced Energy Engineering Science, Interdisciplinary Graduate School of Engineering Science, Kyushu University, Kasuga, Fukuoka 816-8580, Japan

2. Atomic Energy Research Establishment, Bangladesh Atomic Energy Commission, Savar, Dhaka 1349, Bangladesh

E-mail: *khalid.baec@gmail.com, **hashi@nucl.kyushu-u.ac.jp

## ABSTRACT

Hydrogen release behavior from mixed-perovskite oxide, yttrium and cobalt doped barium-zirconate, $BaZr_{0.955}Y_{0.03}Co_{0.015}O_{3-\alpha}$ (BZYC) exposed to $D_2$ or $D_2O$ atmosphere was studied by thermal desorption spectroscopy (TDS) detecting deuterium gas ($D_2$) and heavy water vapor ($D_2O$) separately. The release of hydrogen appeared at around 1000 K. Hydrogen solubility was higher in the BZYC sample exposed to the $D_2O$.

## KEY WORDS

Barium-zirconate, perovskite oxide, hydrogen dissolution, hydrogen solubility, TDS.

## 1. INTRODUCTION

Proton-conducting oxides are promising materials for electro-chemical devices like fuel cell, hydrogen pump, hydrogen sensor, and for nuclear fusion reactor's tritium purification and recovery system. Therefore, the study of hydrogen behavior in such oxide materials is very fundamental and important. Yttrium (Y) and cobalt (Co) doped barium-zirconate ($BaZrO_3$) is a proton conducting oxide has mixed-perovskite structure [1]. In this study, the dissolution of hydrogen in BZYC specimens were done using $D_2$ gas and $D_2O$ vapor separately. Then the hydrogen release and hydrogen solubility were investigated by thermal desorption spectroscopy (TDS) method.

## 2. EXPERIMENTAL

*2.1 Sample preparation*
The disc shape specimen (diameter: 7.7mm, thickness: 2.2mm) was prepared using conventional powder sintering method at 1923 K for 20h, which is described elsewhere [2]. The SEM and EDX mapping were performed before and after polishing the sample to understand its surface properties. From Fig. 1 it is clear that all chemical components of the specimen were uniformly distributed throughout the surface after polishing. The sintered disc specimen was cut using a diamond saw into parallelepiped pieces (7.7 x 2.2 x 0.5 mm$^3$) for $D_2$ gas or $D_2O$ vapor exposure and TDS study.

*2.2 $D_2$ and $D_2O$ exposure*
The cut sample was placed in a quartz glass tube and evacuated to a low pressure (~10$^{-6}$ Pa). At a temperature of 1273 K, vacuum annealing was performed for 30 min to remove the volatile substances from the sample. Then at 873 K, deuterium gas ($D_2$, 1.3 kPa) was exposed to the sample for 1 hr using the apparatus shown in Fig. 2(a) [2]. After the exposure, the quartz glass tube was quenched with room temperature water. Finally, the sample was taken out from the quartz glass tube after recovering the $D_2$ gas. The same sample was used repeatedly for heavy water vapor ($D_2O$, 1.3kPa) exposure and the TDS study.

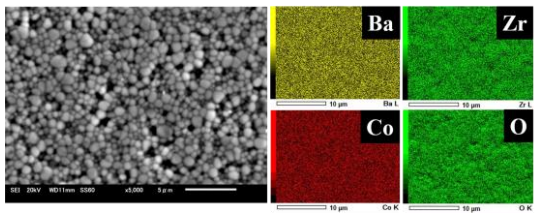

Fig. 1: SEM micrograph and its EDS mapping for the sintered disc sample surface after polishing.

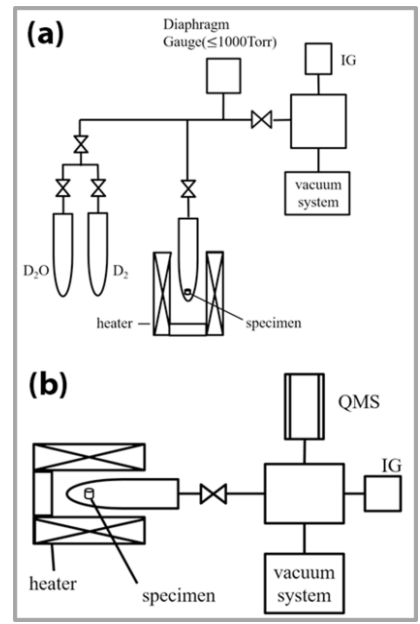

Fig. 2: (a) Schematic of $D_2$ gas and $D_2O$ vapor exposure apparatus, (b) schematic of temperature programmed desorption (TDS) system.

*2.3 $D_2$ and $D_2O$ measurement*

The TDS apparatus used in this experiment to measure the released hydrogen ($D_2$, HD, $D_2O$ and HDO) from the $D_2$ gas and $D_2O$ vapor exposed sample is shown in Fig. 2(b) [2]. First, the baking of the quartz glass tube without the sample was performed with an electric furnace to release and remove substances accumulated in the device at 1273 K for 30 min. Then after cooling down the glass tube to room temperature, the $D_2$ gas exposed sample was placed in the glass tube for TDS study. Then after evacuating the glass tube to a low pressure (~$10^{-6}$ Pa), the temperature of the sample was allowed to raise (at a heating rate of 0.5 K/s) up to 1273 K. After 30 min stay at 1273 K, the heater power was disconnected automatically by using a programable temperature controller. The gas released from the samples was measured using a quadrupole mass spectrometer (Q-MAS) (TSPTT200, INFICON) [3] during this sample heating and cooling down cycle. The above same procedure was also applied for the $D_2O$ exposed sample.

## 3. RESULTS AND DISCUSSION

The ion currents obtained from Q-MAS data for the $D_2$ gas and $D_2O$ vapor exposed samples are shown in Fig 3. It is clear from the figure that in both cases of exposures, release of hydrogen gas ($D_2$ and HD) and water vapor ($D_2O$ and HDO) was observed at around 1000 K. $D_2$ gas was released with an extra peak at 700 K in the case of $D_2$ exposed sample and at 850 K for $D_2O$ exposed sample. Highest release peak was detected as HD gas from $D_2$ exposed sample and there may appear a trend for the hydrogen release as HD > HDO > $D_2$ > $D_2O$. On the other hand, the highest peak was observed as HDO vapor from $D_2O$ exposed sample and there may appear a trend for the hydrogen release as HDO > $D_2O$ > HD > $D_2$.

Table 1: D amount released from $D_2$ and $D_2O$ exposed samples

| Deuterium (D) amount | $D_2$ exposed sample | $D_2O$ exposed sample |
|---|---|---|
| D from $D_2$ and HD (D/M) | 4.6 x $10^{-4}$ | 4.5 x $10^{-4}$ |
| D from $D_2O$ and HDO (D/M) | 4.4 x $10^{-4}$ | 1.3 x $10^{-3}$ |
| Total D (D/M) | 9.0 x $10^{-4}$ | 1.8 x $10^{-3}$ |
| Fraction of deuterium gas (%) | 51 | 25 |
| Fraction of water vapor (%) | 49 | 75 |

The total amounts of D-containing gas released from the $D_2$ or $D_2O$ exposed samples and its ratios are listed in Table 1. Total amount of deuterium (D) released from $D_2$ exposed sample was 9.0 x $10^{-4}$ (D/M), where released deuterium gas (~ 4.6 x $10^{-4}$ D/M) and water vapor (~ 4.4 x $10^{-4}$ D/M) were contributed equally. On the other hand, total amount of D released from $D_2O$ exposed sample was 1.8 x $10^{-4}$ (D/M), where three-fourth of the released D was water vapor (~ 1.3 x $10^{-4}$ D/M). It is also clear that total amount of deuterium released from $D_2O$ exposed sample is twice than the $D_2$ exposed sample. Percentage of water vapor released from the $D_2O$ exposed sample is higher than that of the $D_2$ exposed sample, whereas an opposite result appears for deuterium gas release. The obtained result of the experiment suggests that, under $D_2O$ exposed condition and at 1000 K operating temperature, Y and Co doped barium-zirconate may preferable for electrochemical application due to its higher hydrogen solubility.

## 4. CONCLUSIONS

In this study, hydrogen (deuterium) was dissolved in $BaZr_{0.955}Y_{0.03}Co_{0.015}O_{3-\alpha}$ by the exposure of deuterium gas or heavy water vapor, and the amount of dissolved deuterium in the sample was measured by TDS method. It was observed that deuterium solubility is higher in the BZYC sample under the $D_2O$ exposure condition and release of deuterium was appeared at around 1000 K.

## ACKNOWLEDGEMENTS
We would like to thank our laboratory members for their cordial support to operate the experimental instruments.

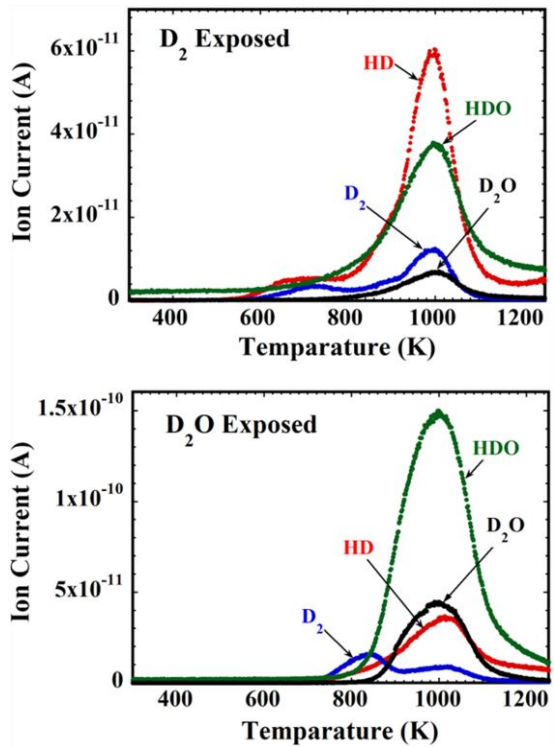

Fig. 3: TDS results for (a) $D_2$ gas, and (b) $D_2O$ vapor exposure samples.